\documentclass[12pt]{article}
\usepackage{epsfig,psfig,graphicx,rotating}
\def\be{\begin{equation}}
\def\ee{\end{equation}}
\begin{document}
\centerline{\bf SELF-GRAVITATING PHASE TRANSITIONS:} 
\centerline{\bf POINT PARTICLES, BLACK HOLES AND STRINGS}

\begin{center}{\bf Norma G. Sanchez}
\footnote{Norma.Sanchez@obspm.fr}
\end{center}
\begin{center}{Observatoire de Paris, LERMA, Laboratoire Associ\'e au CNRS UMR 8112, 61, Avenue de l'Observatoire, 75014 Paris, France.}
\end{center}

\centerline{\bf Abstract}

We compute the quantum string entropy $S_s(m,j)$ of the microscopic string states of mass $m$ and spin $j$ in two physically relevant backgrounds: Kerr  (rotating) black holes and de Sitter (dS) space-time. We find a {\bf new} formula for the quantum gravitational entropy $S_{sem} (M, J)$, as a function of the usual Bekenstein-Hawking entropy $S_{sem}^{(0)}(M, J)$. We compute the quantum string emission by a black hole in de Sitter space-time (bhdS). In {\bf all} these cases: (i) strings with the highest spin, and (ii) in dS space-time, (iii) quantum rotating black holes, (iv) quantum dS regime, (v) late bhdS evaporation, we find a {\bf new} gravitational phase transition with a common distinctive {\bf universal} feature: A square root {\bf branch point} singularity in any space-time dimensions.  This is the same behavior as for the thermal self-gravitating gas of point particles (de Vega-Sanchez transition), thus describing a new {\bf universality class}.

Nous calculons l'entropie $S_s(m, j)$ des \'etats microscopiques de masse $m$ et spin $j$ des cordes quantiques dans deux espaces-temps physiquement relevants: Le trou noir en rotation (de Kerr) et l'espace-temps de de Sitter (dS). Nous trouvons une {\bf nouvelle} formule pour l'entropie gravitationnelle quantique $S_{sem} (M, J)$  comme fonction de l'entropie de Bekenstein-Hawking $S_{sem}^{(0)}(M)$. Nous calculons l' \'emission quantique des cordes par un trou noir avec constante cosmologique (bhdS). Dans {\bf tous} ces cas: (i) cordes avec le spin maximal, et (ii) dans l'espace-time de dS,  (iii) trous noirs de grand moment angulaire, (iv) regime quantique de dS, (v) derni\`ere \'etape d'\'evaporation bhdS, nous trouvons une {\bf nouvelle} transition de phase avec la m\^eme caract\'eristique distinctive {\bf universelle} : un {\bf point de ramification} racine carr\'ee \`a la transition, pour toute dimension de l'espace-temps. C'est le m\^eme comportement  que pour le gaz auto gravitant de particules ponctuelles (transition de de Vega-Sanchez), d\'efinissant ainsi une nouvelle {\bf classe d'universalit\'e}.
\\

A theory of quantum gravity, such as string theory, should account for an unified and consistent description of both black holes and elementary particles, and the physics of the early universe as well. \\
A central object in string theory is $\rho_s (m,j)$, the microscopic string density of states of mass $m$ and spin $j$. We find $\rho_s (m,j)$ by solving the non-linear and quantum string {\bf dynamics} in the curved space-times considered, refs. \cite {1}-\cite{5}. We compute the quantum string entropy from the microscopic density of states $\rho_s (m,j)$ in two physically relevant  space-times: the rotating (Kerr) black hole and the de Sitter space-time, refs \cite{6},\cite{7}. Combination of the two effects at the transition: black hole and cosmological constant are also treated for a black hole in de Sitter space-time (bhdS), ref. \cite{7}. 
\\
The Hawking temperature $T_{sem}$ and the string (Hagedorn) temperature $T_s$ are the same concept in different (semiclassical and quantum) gravity regimes,  ref. \cite{2}. Similarly, it holds for the Bekenstein-Hawking entropy  and  the string entropy, ref. \cite{2}. $T_s$ is the precise quantum dual of $T_{sem}=\hbar c /(2\pi k_B L_{c\ell})$, $ L_{c\ell}$ being the classical radius: $2R_S/(D-3)$, ($R_S$ is the black hole radius), or $c/H$ for Sitter space-time, $H$ being the Hubble constant, (in any $D$ space-time dimensions).  For $L_{c\ell}\gg \ell_{Planck}$, ie. for low curvature or low $H\ll c/\ell_{Planck}$, (semiclassical gravity regime), the Bekenstein-Hawking entropy $S_{sem}^{(0)}$ is the leading term of the whole entropy $S_{sem}$, {\bf but} for high curvature or very small radius near $\ell_{Planck}$, (quantum gravity regime),  a {\bf new} phase transition operates and the whole entropy $S_{sem}$ is drastically  {\bf different} from the Bekenstein-Hawking entropy $S_{sem}^{(0)}$, refs \cite{6},\cite{7}. 
\\ 
The phase transitions we found for quantum strings in Minkowski, Kerr and de Sitter backgrounds, as well as for the quantum regimes of rotating black holes and de Sitter space, are {\bf all} of the same nature: a {\bf square root} branch point singularity, (the value of the critical temperature being different in each case). This phase transition does not occur in Anti de Sitter space-time (AdS): in the AdS background, the density of mass states is finite, the negative cosmological constant "pushes" the critical temperature to infinity and there is no singularity at all, refs \cite{1}, \cite{2}.
\\
The origin of this string or gravitational {\bf square root} phase transition is gravity at finite temperature or in the presence of a repulsive force (angular momentum, positive cosmological constant). This behavior is similar to that found for the selfgravitating gas of point particles (de Vega-Sanchez transition), refs.\cite{8}, \cite{9}. In our context here, the temperature is not an external parameter, but {\bf intrinsic} to the theory, determined by the mass of the system (strings, or space-time backgrounds). Furthemore, our treatment here is not statistical mechanical , but {\bf microscopic} or dynamical.   
\\
Here, we focalizes on the phase transition we found for strings and the quantum black hole and de Sitter regimes. More results and implications on these new issues can be found in refs. \cite{6},\cite{7}. Black hole evaporation ends as quantum string decay into elementary particles (most massless), ie pure (non mixed) quantum states refs \cite {2},\cite{6}, \cite{7}.\\
{\bf Quantum String Entropy in the Kerr background}: The string entropy $S_s(m,j)$ in a Kerr black hole background is given by
$\rho_s(m,j)= e^{\frac{S_s(m,j)}{k_B}} $, where the microscopic string density of states of mass $m$ and spin $j$, $\rho_s(m, j)$, has been found in ref. \cite{6}  and can be expressed as
$$
\rho_s(m, j) \sim \Big(\frac{S_s^{(0)}}{k_B}\Big)^{-a} ~ e^{\Big(\frac{S_s^{(0)}}{k_B}\Big)} ~F(S_s^{(0)}, j)~~, 
$$
$$  F(S_s^{(0)}, j) = \Delta_s^{-a-1} e^{\frac{S_s^{(0)}}{k_B} \frac{(1 - \Delta_s) ^2}{2 \Delta_s}} 
\cosh^{-2} \Bigg(  \frac{S_s^{(0)}}{4 k_B} \frac{(1 - \Delta_s^2)}{\Delta_s} \Bigg) 
$$
~~~~~~~~~~$\Delta_s = \sqrt{1 - \frac{4 j }{m^2 \alpha^{'} c}}=\sqrt{1 - \frac{j}{\hbar} \Big(  \frac{k_B~b}{S_{s}^{(0)}} \Big)^2}~~,~~~4j\leq m^2 \alpha^{'} c$~. 

$S_s^{(0)}$ is the zero order string entropy for $j=0$:   $S_s^{(0)} =   \frac{1}{2} \frac{mc^2}{t_s}$ ~~~(closed~ strings), ~~~~
$S_s^{(0)} =  \frac{mc^2}{t_s}$~~~(open~ strings), and $t_s$ is the string  Hagedorn temperature, ~~~~$t_s =   \frac{m_s c^2}{k_B b}$, ~~~being ~~$b = 2 \pi \sqrt{ \frac{D- 2 }{6}}$, $a = D$  (closed strings),~~   $a = (D-1)/2$ (open strings). $D$ is the number of space-time dimensions. $m_s$ is the fundamental string mass: $m_s = \sqrt{\frac{\hbar}{\alpha'c}}\equiv~\frac{\ell_s}{\alpha'}$. $\alpha'$ is the fundamental string constant ($\alpha'^{-1}$ is a mass linear density), $\ell_{s}$ is the fundamental string length. $\Delta_s$ can be expressed as :
$$
\Delta_s= \sqrt{1 - \frac{j}{\hbar} ~\Big(\frac{m_s}{m} \Big)^2} =  \sqrt{1 - \frac{j}{\hbar}~ \Big(\frac {t_s}{T} \Big)^2} ~~,~~ T = \frac{m c^2}{k_B b}.
$$
$F({m}, j)$ takes into account the effect of the angular modes $j$, being $ F({m},j=0) = 1$. Therefore, the string entropy $S_s(m,j)$ in the Kerr background is given by
$$
S_s(m,j) = S_s^{(0)} - a ~k_B~\ln \Big (~\frac{S_s^{(0)}}{k_B}~ \Big)~+~k_B~\ln F(S_s^{(0)}, j)
$$
The last new term $k_B\ln F(S_s^{(0)},j)$ is enterely due to the spin $j \neq 0$. Interestingly enough, this formula expresses the string  entropy $S_s(m,j)$ for mass $m$ and spin $j$ in terms of the string entropy $S_s^{(0)}$ for $j=0$. For $j=0$, we recover the  usual expression. The logarithmic terms have a negative sign, the effect of the spin is to reduce the entropy. $S_s(m,j)$ is maximal for $j=0$ (ie, for $\Delta_s = 1$).\\
For $j < (m/m_s)^2$, and $m \gg m_{s}$, that is for low $j$ and very excited string states, $S_s^{(0)}(m,j)$ is the leading term, but for high $j$, that is $j \rightarrow m^2 \alpha^{'} c$, ie $\Delta_s \rightarrow 0$, the situation is {\it very different} as we see it below. Moreover, $S_s(m,j)$ allows us to write the full expression for the gravitational Kerr black hole entropy $S_{sem}$, as a function of the Bekenstein-Hawking entropy $S_{sem}^{(0)}$.\\
{\bf Extremal String States. A New Transition}: In ref. \cite {6} we considered a new kind of string states, the states in which $j$ reachs its {\bf maximal} value, that is $j =  m^2 \alpha^{'} c$, we call these states {\bf "extremal string states"}. In this case, $\Delta_s = 0$. For $\Delta_s \rightarrow 0$, the entropy $S_s(m,j)$ behaves as
$$
S_s(m,j)_{extremal} =   -(a + 1)~k_B~ \ln \frac {\Delta_s}{2}~ +~ k_B~\ln 2~ +~ \Delta_s ~\Big(~\frac{3} {4}~S_s^{(0)}~-~ak_B \Big)~+ ~ O (\Delta_s^{2})
$$
In the extreme limit $(j/\hbar) \rightarrow (m/m_s)^2 $:~~$\Delta_{s              ~extremal}= \sqrt{2}\sqrt{1 - \Big(\frac{j}{\hbar}\Big)^{1/2} ~\frac{m_s}{m}}$\\ and $S_s(m,j)_{extremal}$ is dominated by
$$
S_s(m,j)_{extremal} =  -(a + 1)~k_B~ \ln ~ \sqrt{1 - \Big(\frac{j}{\hbar}\Big)^{1/2} ~\frac{t_s}{T}}~~+~O(1)
$$
This shows a {\bf phase transition} takes place at $T \rightarrow \sqrt{(j/\hbar)}~t_s$, we call it {\bf extremal} transition. This is {\bf not} the usual (Hagedorn/Carlitz) string phase transition occuring for $m \rightarrow \infty $, $T \rightarrow t_s$; such transition is also present for $j\neq 0$ since $\rho (m,j)$ has the same $m \rightarrow \infty $ behaviour as $\rho(m)$. The extremal transition we found here is a {\bf gravitational} like phase transition: the square root {\bf branch point} behaviour near the transition is analogous to that found in the thermal self-gravitating gas of (non-relativistic) particles (by both mean field and Monte Carlo methods), refs. \cite {8},\cite{9}. And this is also the same behaviour found for the microscopic density of states and entropy of strings in de Sitter background, refs. \cite{2}, \cite{7}.\\
As pointed out in ref.\cite {2}, this string behaviour is {\bf universal}: the logarithmic singularity in the entropy (or pole singularity in the specific heat) holds in any number of dimensions, and is due to the gravitational interaction in the presence of temperature, similar to the Jeans's like instability at finite temperature but with a more complex structure.
A particular new aspect here is that the transition shows up at high angular momentum, (while in the thermal gravitational gaz or for strings in de Sitter space, angular momentum was not considered, (although it could be taken into account)). 
Since $j\neq 0$, the extremal transition occurs at a temperature  $ T_{sj}~=~ \sqrt{j/\hbar}~T_s $,  {\bf higher} than the string temperature $T_{s}$.
That is, the angular momentum which acts in the sense of the string tension, appears in the transition as an "effective string tension" $(\alpha^{'}_j~)^{-1}$: a smaller  $\alpha^{'}_j~  \equiv ~ \sqrt{\hbar/j}~ \alpha^{'}$ ~ (and thus a {\bf higher} tension).\\     
{\bf Quantum String Entropy in de Sitter Background}: The entropy of quantum strings in de Sitter background is defined by
$ \rho_{s}(m, H) = e^{\frac{S_{s}(m,H)}{k_B}}$, 
where $\rho_{s}(m, H)$  is the microscopic string density states of mass $m$ in de Sitter space time, and $H$ is the Hubble constant. In terms of the zero order  string entropy in flat space time ($H=0$) $S_s^{(0)}$,
the string density of mass states $\rho_s(m, H)$ for both open and closed strings can be expressed as :$$\rho_s (m, H) =\Big(  \frac{S_s^{(0)}}{k_B} \sqrt{f(x)} \Big)^{-a}~       e^{\Big(\frac{S_s^{(0)}}{k_B}\sqrt{f(x)}\Big)}~F(x)  
~~~~,~~~~ F(x) = \frac{1}{\Delta_s \sqrt{f(x)}},$$ ~~~~
$$f(x)= \frac{2}{1+\Delta_s}~~, ~~a=\frac{(D-1)}{2} ~~(open),~~ a=D ~~(closed) , $$ 
$x$~  being the dimensionless variable~~$x(m, H)\equiv \frac{1}{2}\Big(\frac{m}{M_s}\Big)= 
\frac{m_s}{b M_s}\frac{S_s^{(0)}}{k_B}$~~ and ~~$\Delta_s$~~ is given by:
$$\Delta_s~~=~~\sqrt{1-4x^2}~~=~~\sqrt{1 -\Big(\frac{m}{M_s}\Big)^2}$$ 
$M_s$ is the highest string mass in de Sitter space time ~\cite{1},\cite{2},\cite{3},\cite{5}$$
M_s = \frac{L_{c\ell}}{\alpha'} = \frac{c}{H ~ \alpha'};~~~~\Bigg(\frac{m_s}{M_s} \Bigg) = \frac{\ell_{s}}{L_{c\ell}}= \frac{H}{c}\ell_{s}$$
$L_{c\ell}= c/H$ being the classical de Sitter radius. Furthermore, $M_s$ defines the quantum string de Sitter length $L_s$ and the de Sitter string temperature $T_s$: 
$$
L_s = \frac{\hbar }{M_s ~c}= \frac{\ell_s^2}{L_{c\ell}}=\frac{\hbar\alpha'}{c^2}H~~,~~
T_s = \frac{1}{2\pi k_B} ~ M_s~c^2 = \frac{\hbar c}{2 \pi k_B} ~\frac{1}{L_s}= \frac{1}{2 \pi k_B} ~\frac{c^3}{H \alpha'}
$$
$T_s$ appears to be a true critical temperature as we will see below. \\
For small $x$, (small $H m\alpha '/c$), $f(x)$ can be naturally expressed as a power expansion in $x$. In particular, for $H=0$, we have $x=0$ and $f(x)=1$, and we recover the flat space time string solution: $
 \rho_s(m, H=0) \simeq  \Big( \frac{S_s^{(0)}}{k_B} \Big)^{-a} ~e^{\big( \frac{S_s^{(0)}}{k_B} \big)}$.
From $\rho_s(m,H)$, we can read the full string entropy in de Sitter space :
$$
S_s(m,H) = \hat {S_s}^{(0)}(m,H) 
-a~k_B~\ln ~\big(\frac{\hat {S_s}^{(0)}(m, H)}{k_B}\big) + k_B ~\ln F(m,H)
$$
$$\hat {S_s}^{(0)}(m,H)\equiv S_s^{(0)}\sqrt{f(x)} $$
The mass domain is  $ 0 \leq  m \leq M_{s}$, ie. $ 0 \leq x \leq 1/2 $,  (which implies $   0\leq \Delta_s  \leq  1$,  ie. $1 \leq f(x)\leq 2 $). All terms in the entropy $S_s(m,H)$ except the first one have negative sign.  For $\Delta_s \neq 0$, (ie. $ m \neq M_{s}$),  the entropy $S_s(m,H)$ of string states in de Sitter space is smaller than the string entropy for $H=0$. The effect of the Hubble constant is to reduce the entropy.\\
For low masses  $m \ll M_{s}$, the entropy is a series expansion in $(H m\alpha' /c  \ll 1)$, like a low H expansion around the flat $H=0$ solution, $S_s^{(0)}$ being its leading term. But for high masses  $m \rightarrow M_s$, that is $(H m \alpha^{'}/c) \rightarrow 1$, (i.e. $\Delta_s \rightarrow 0$), the situation is {\bf very different} as we see it below. Moreover, $S_s(m,H)$ allows us to write the whole expression for the semiclassical de Sitter entropy $S_{sem}(H)$, as a function of the (Bekenstein-Hawking) de Sitter entropy $S_{sem}^{(0)}(H)$.\\
{\bf The String de Sitter Phase Transition}: For high masses $m \sim M_s$, (or in terms of temperature $T \sim T_s$), the entropy behaves as:
$$
S_s(T,H)_{T \sim T_s} = k_B \ln \sqrt{1 - \frac{T}{T_s}}~-k_B\ln~2 ~+~k_B \frac{ b}{\sqrt{2}}~(\frac{T_s}{t_s})~-~ a k_B \ln~(\frac{T_s}{t_s})
$$
where ~ $T = m c^2/2 \pi k_B $. We see that a phase transition takes place at $m = M_s$, ie,  $T = T_s$. This is a {\bf gravitational} like phase transition: the square root {\bf branch point} behaviour near the transition is analogous to the thermal self-gravitating gas phase transition of point particles, refs. \cite {8},\cite{9}. This is also the same behaviour of the microscopic density of states and entropy of strings with the spin modes included ref. \cite{6}. \\
The transition occurs at the string de Sitter temperature $T_{s}$ {\bf higher} than the (flat space) string temperature $t_{s}$:  $(T_s/t_s)=  (b M_s /2\pi m_s)  = (b L_{c\ell}/2\pi \ell_s)$. 
This is so since in de Sitter background, the flat space-time string mass $m_s$, (or Hagedorn temperature $t_s$) is the scale mass, (or scale temperature), in the {\bf low} $Hm$ regime. For high masses, the critical string mass, (temperature), in de Sitter background is $T_s$, instead of $m_s$, $(t_s)$.  In de Sitter space, $H$ "pushes" the string temperature beyond its flat space Hagedorn value $t_s$ . That is, $H$, which acts in the sense of the string tension, does appear in the transition temperature as an "effective string tension" $(\alpha^{'}_H) ^{-1} $ : a smaller  $\alpha^{'}_H = (\hbar/c)(~H \alpha'/c)^2$, (and thus a {\bf higher} tension). The effect of $H$ in the transition is similar to the effect of angular momentum $j$. \\
Furthemore, for high masses ($m \sim M_s$), the {\bf partition function}  $\ln Z$  of  a {\bf gas of strings} in de Sitter space-time,  behaves as 
$$
(\ln Z)_{T \sim T_s}\sim V_{D-1}\left(\frac{k_B T_{sem}}{\hbar c}\right)^{D-1}~~
 \sqrt{1 - \frac{T_{sem}}{T_s}} 
$$
where $T_s$ is the string de Sitter temperature and $ T_{sem}$ is the semiclassical (Gibbons-Hawking) de Sitter temperature. $\ln Z$  shows a singular behavior for $T_{sem} \rightarrow  T_{s}$ which is general for {\bf any} space-time dimensions $D$. Again,  this is a {\bf square root branch point} at $T_{sem}= T_s$.  That is, a {\bf phase transition} takes place for $T_{sem} \rightarrow T_s$, which implies $M_{c\ell}\rightarrow  m_{s}$,  $L_{c\ell}\rightarrow \ell_{s}$.\\ 
Furthemore, the high mass behaviour of $\ln Z$ implies that $T_{sem}$ has to be bounded by $T_s$,  $(T_{sem} < T_s)$, which means: $L_{c\ell} > \ell_s, ~~i.e., ~~H < {c}/{\ell_s}$.  In the de Sitter string phase transition  $T_{sem} \rightarrow T_{s}$, $H$ reachs a maximum value sustained by the string tension $\alpha'^{-1}$ (and the fundamental constants $\hbar$, $c$ as well):
$$H_s = c ~\sqrt{\frac{c}{\alpha'\hbar}}, ~~~~ i.e.,~~\Lambda_s = \frac{1}{2 {\ell_s}^2}(D-1)(D-2)$$
The highly excited $m  \rightarrow M_s$ string gas in de Sitter space undergoes a phase transition at high temperature $T_{sem} \rightarrow T_{s}$, into a condensate stringy state. This means that the background itself becames a string state. \\
These results  also allow to consider the string regimes of a black hole in a de Sitter (or asymptotically) de Sitter background. This allow to study the effects of the cosmological constant $\Lambda$ on the quantum string emission by black holes, and the string bounds on the semiclassical (Gibbons-Hawking) {\bf black hole-de Sitter} (bhdS) temperature~ $T_{sem~bhdS}$, ref. \cite{7}. We computed the quantum string emission cross section $\sigma_{string}$ by a black hole in de Sitter (or asymptotically de Sitter) space-time (bhdS), ref. \cite{7}. For $T_{sem~bhdS}\ll T_{s}$, (early evaporation stage), it shows the QFT Hawking emission with temperature $T_{sem~bhdS}$, (semiclassical regime). For $T_{sem~bhdS}\rightarrow T_{s}$, $\sigma_{string}$~exhibits a phase transition into a  string de Sitter state of size $L_s = \ell_s^2/L_{c\ell}$, ($\ell_s= \sqrt{\hbar \alpha'/c}$), and string de Sitter temperature $T_s$. For high masses ($m \sim M_s$), we find for the $\sigma_{string}$ leading behavior: 
$$
\sigma_{string} ~~(T \sim T_s) \sim V_{D-1}~\Gamma_A~\left(\frac{k_B T_{sem~bhdS}}{\hbar c}\right)^{D-1}
\sqrt{1~-~\frac{T_{sem~bhdS}}{T_s}}
$$
Instead of featuring a single pole singularity in the temperature (Carlitz transition), it features a square root {\bf branch point} (de Vega-Sanchez transition, refs.\cite{8},\cite{9}). {\bf New} bounds on the black hole radius $r_g$~ emerge in the  bhdS string regime: it can become $r_g = L_s/2$, or it can reach a more quantum value, $r_g = 0.365 ~\ell_s$, ref. \cite{7}.\\
{\bf Semi-classical and quantum (string) black hole regimes}:
Our analysis of the string canonical partition function, the black hole quantum string emission and the string bounds on the black hole, shows that a semiclassical black hole (BH) with size $L_{c\ell}$, mass $M$, temperature $ T_{sem}$, density of states $ \rho_{sem}$ and entropy $S_{sem}$, namely $(BH)_{sem}= (L_{c\ell}, M,  T_{sem}, \rho_{sem}, S_{sem})$, evolves through evaporation into a quantum string state of size $L_{s}$, mass $m$, temperature $T_{s}$, density of states  $\rho_{s}$ and entropy $S_{s}$, namely $(BH)_{s}$ = $(L_{s}, m , T_{s}, \rho_{s}, S_{s})$. The quantities in the set $(BH)_{sem}$ are precisely the semiclassical expressions of the respective ones in the set $(BH)_{s}$. In the quantum string regime, the black hole size $L_{c\ell}$ becomes the string size $L_{s}$, the Hawking temperature $T_{sem}$ 
becomes the string temperature $T_s$, the black hole entropy $S_{sem}$ becomes the string entropy $S_s$. The sets $(BH)_{s}$ and $(BH)_{sem}$ are the same quantities but in different (quantum and semiclassical/classical) domains. That is, $(BH)_{s}$ is the quantum dual of the semiclassical set $(BH)_{sem}$ in the precise sense of the wave-particle (de Broglie) duality. This is the usual classical/quantum duality but in the gravity domain, which is {\bf universal}, not linked to any symmetry or isommetry nor to the number or the kind of space-time dimensions. 
From the semiclassical and quantum (string) black hole regimes $(BH)_{sem}$ and $(BH)_{s}$, we can write the full gravity entropy $S_{sem}(M, J)$ for the {\bf Kerr black hole} such that it becomes the string entropy $S_s(m,j)$ in the quantum string regime, namely:
$$
S_{sem}(M, J) = S_{sem}^{(0)}(M, J) - a ~k_B~\ln \Big (~\frac{S_{sem}^{(0)}(J, M)}{k_B}~ \Big)~+~k_B~\ln~ F(S_{sem}^{(0)}, J)           
$$
where $S_{sem}^{(0)}(M, J)$ is the Kerr black hole Bekenstein-Hawking entropy and $F(S_{sem}^{(0)}, J)$ is given by
$$
F = \Delta^{-1} \Big(\frac{1 + \Delta}{2\Delta}\Big)^{a}~e^{\Big(\frac{1 -\Delta}{1 +\Delta}\Big) \frac{S_{sem}^{(0)}(M, J)}{\Delta k_B}} 
\cosh^{-2} \Bigg(\frac{(1 - \Delta)}{\Delta}\frac{S_{sem}^{(0)}(M, J)}{2 k_B}\Bigg)
$$ 
For $J=0:~~~~~~F = 1 ~~~~, ~~~~ S_{sem}^{(0)}(M,J=0)~\equiv~ S_{sem}^{(0)} ~=~ 4 \pi k_{B} \Big(\frac{M}{m_{Pl}}\Big)^2$

$S_{sem}^{(0)}$ being the Schwarzschild black hole Bekenstein-Hawking entropy. $\Delta$ is given by :
$$
\Delta = \sqrt{1 - \Big(\frac{J}{\hbar}\Big)^2 \Big(\frac{4 \pi k_B~}{S_{sem}^{(0)}} \Big)^2}=~ \sqrt{1 - \Big(\frac{J}{\hbar}\Big)^2 \Big(\frac{m_{Pl}}{M} \Big)^4}~=~ \sqrt{1 - \Big(\frac{J}{\hbar}\Big)^2 \Big(\frac{T_{sem}}{T} \Big)^2} 
$$
where $T = M c^2 / 8\pi k_B$ and  $T_{sem}$ is the Schwarzshild Hawking temperature. $m_{Pl}$ is the Planck mass. Notice the  {\bf new} term  $k_B\ln F(S_{sem}^{(0)},J)$ in $S_{sem}(M, J)$ enterely due to $J \neq 0$.
For $\Delta \neq0$, the effect of the angular momentum is to reduce the entropy. $S_{sem}(M, J)$ is maximal for $J=0$ (ie, for $\Delta = 1$); for $ J=0$ we have:  $S_{sem}(M)~ = ~S_{sem}^{(0)}~-~ a ~k_B~\ln S_{sem}^{(0)}$.\\
$S_{sem}(M,J)$ provides the full Kerr black hole entropy as a function of the Kerr Bekenstein-Hawking entropy $S_{sem}^{(0)}(M,J)$. Interestingly enough, the full Kerr black hole entropy $S_{sem}(M, J)$ can be also written enterely in terms of the Schwarschild Bekenstein-Hawking entropy $S_{sem}^{(0)}$, this is done in ref. \cite{6}. For $M\gg m_{Pl}$  and  $J < GM^2/c$,   $S_{sem}^{(0)}$   is the leading term of this expression, {\bf but} for high angular momentum, (nearly extremal or extremal case $J= GM^2/c$), a gravitational {\bf phase transition} operates and the whole entropy $S_{sem}$ is drastically  {\bf different} from the Bekenstein-Hawking entropy $S_{sem}^{(0)}$, as we precisely see it below.\\
{\bf A New feature. The Extremal Black Hole Phase Transition:}
When $J$ reachs its {\bf maximal} value, that is $J = M^2 G/c^2$, then $\Delta = 0$ and the term  $S_{sem}^{(0)}(M,J)$ is minimal :
$S_{sem}^{(0)} (M,J)_{extremal} ~ = ~\frac{1}{2}~S_{sem}^{(0)}$.  
But for $\Delta \rightarrow 0$, the last term in the expression for $S_{sem}(M,J)$ substracts the first one, the poles in $\Delta$ cancel out, yielding:
$$
S_{sem}(M,J)_{extremal} =   -(a + 1)~k_B~ \ln \frac {\Delta}{2}~ +~ k_B~\ln 2~ +~ \Delta ~\Big(~\frac{3} {4}~S_{sem}^{(0)}~-~ak_B~\Big)~+ ~ O(\Delta^{2})
$$
In the extreme limit $(J/\hbar) \rightarrow (M/m_{Pl})^2 $ :~~ $\Delta_{extremal} = ~ 2 ~\sqrt{1 - \Big(\frac{J}{\hbar}\Big)^{1/2}~ \frac {m_{Pl}}{m}}$\\and $S_{sem}(M,J)_{extremal}$ is dominated by
$$ S_{sem}(M,J)_{extremal} =  -(a + 1)~k_B~ \ln ~\sqrt{1 - \Big(\frac{J}{\hbar}\Big)^{1/2}~\frac{t_{Pl}}{T}}~+~O(1)
$$ This shows that a {\bf phase transition} takes place at $T \rightarrow \sqrt{(J/\hbar)}~t_{Pl} $, equivalently , at $T \rightarrow (J/\hbar)~T_{sem} $,  we call it {\bf extremal transition}. $t_{Pl}$ is the Planck temperature. 
The characteristic features of this gravitational transition can be discussed on the lines of the extremal string transition we analysed for the extremal string states. Our discussion on the string case translates into the respective black hole quantities  but we do not extend on these new features and implications here.\\
{\bf Semi-classical and quantum (string) de Sitter regimes}:
 From the microscopic string density of mass states $\rho_s(m, H)$, we have shown that for  $m\rightarrow M_s$,~ i.e. $T\rightarrow T_s$, the string undergoes a phase transition into a semiclassical phase with mass $M_{cl}$ and  temperature $T_{sem}$.  Conversely, from the string canonical partition function in de Sitter space and from the quantum string emission by a black hole in de Sitter space, we have shown that for $T_{sem}\rightarrow T_s$, the semiclassical (Q.F.T) regime with Hawking-Gibbons temperature $T_{sem}$ undergoes a phase transition into a string phase at the string de Sitter temperature $T_{s}$. This means that in the quantum string regime, the semiclassical mass density of states $\rho_{sem}$ and entropy $S_{sem}$ become  the string density of states $\rho_s$ and string entropy $S_{s}$ respectively. Namely, a semiclassical de Sitter state, $(dS)_{sem}= (L_{c\ell}, M_{c\ell},  T_{sem}, \rho_{sem}, S_{sem})$, undergoes a phase transition  into a quantum string state $(dS)_{s}$ = $(L_{s}, m , T_{s}, \rho_{s}, S_{s})$. 
The sets $(dS)_{s}$ and $(dS)_{sem}$ are the same quantities but in different (quantum and semiclassical/classical) regimes. This is the usual classical/quantum duality but in the gravity domain, which is {\bf universal}, not linked to any symmetry or isommetry nor to the number or the kind of dimensions. 
From the semiclassical and quantum de Sitter regimes $(dS)_{sem}$ and $(dS)_{s}$, we can write the full de Sitter entropy $S_{sem}(H)$, with quantum corrections included, such that it becomes the de Sitter string entropy $S_s(m,H)$ in the string regime: the full de Sitter entropy $S_{sem}(H)$ is given by $$ S_{sem}~(H) = \hat{S}_{sem}^{(0)}~(H)
-a~k_B~\ln ~(\frac{\hat{S}_{sem}^{(0)}~(H)}{k_B})+ k_B \ln~F(H)$$
$$\hat{S}_{sem}^{(0)}~(H)\equiv S_{sem}^{(0)}~(H) \sqrt{f(X)}~~~,~~~F(H) = \Delta \sqrt{f(X)}~~,~~ a=D~~,~~
f(X)= \frac{2}{1+\Delta},$$
$$\Delta ~\equiv~\sqrt{1-4X^2}~=~ \sqrt{1 -\Big(\frac{\pi k_B}{S_{sem}^{(0)}(H)}\Big)^2}, ~~X(H)\equiv \frac{\pi k_B}{2 S_{sem}^{(0)}(H)}= \frac {M_{sem}}{M_{cl}}= \Big(\frac{m_{Pl}}{M_{cl}}\Big)^2.   
$$
$S_{sem}^{(0)}(H)$ is the usual Bekenstein-Hawking entropy of de Sitter space. $M_{cl}= c^3/GH $ is the de Sitter mass scale, $M_{sem}= m_{pl}/M_{cl}$ is the semiclassical mass. \\
In this expression, the mass domain is $m_{pl} \leq M_{c\ell} \leq \infty$, that is,  $0\leq X \leq 1/2$. (ie. $0\leq \Delta \leq 1$). The same formula but with  $\hat{X}(H) = S_{sem}^{(0)}(H)/2\pi k_B$, instead of $X(H)$,  describes $S_{sem}(H)$ in the mass domain  $0 \leq M_{cl} \leq m_{pl}$. This provides the whole de Sitter entropy $S_{sem}(H)$ as a function of the Bekenstein-Hawking entropy $S_{sem}^{(0)}(H)$.\\
The limit $X \rightarrow 0$ corresponds to $M_{cl}\gg m_{Pl}$, that is $L_{c\ell}\gg \ell_{Pl}$ , (low $H \ll c/\ell_{Pl}$ or low curvature regime). In this limit, $\Delta \rightarrow 1$,  $f(X)\rightarrow 1$ and  $S_{sem}^{(0)}(H)$  is the leading term of $S_{sem}(H)$, with its logarithmic correction:
$$
S_{sem}(H) = S_{sem}^{(0)}(H) -ak_B~\ln \Big(\frac{S_{sem}^{(0)}(H)}{k_B}\Big) 
$$
But for {\bf high} Hubble constant, $H \sim c/\ell_{Pl}$, (ie. $M_{cl}\sim m_{Pl}$), (that is {\bf high} curvature or quantum gravity regime), $S_{sem}^{(0)}(H)$ is sub-dominant, a gravitational {\bf phase transition} operates and the whole entropy $S_{sem}(H)$ is drastically  {\bf different} from the Bekenstein-Hawking entropy $S_{sem}^{(0)}(H)$, as we precisely see below.\\ 
{\bf The de Sitter Gravitational Phase Transition} : For $\Delta \rightarrow 0$, that is for $ M_{c\ell}\rightarrow M_{sem}$, the  full de Sitter entropy $S_{sem}(H)$ behaves as:$$
S_{sem}(H)_{\Delta \sim 0}~~ =~~ k_B~ \ln \Delta ~+~O(1)$$
In this limit, the Bekenstein-Hawking entropy $S_{sem}^{(0)}(H)$ is sub-leading, O(1), (~$S_{sem}^{(0)}(H)_{\Delta = 0}  = \pi k_B$~). In terms of the mass, or temperature:$$
\Delta~ =~ \sqrt{1 -  \Big(\frac{m_{Pl}}{M_{cl}} \Big)^4}~=~ \sqrt{1 - \Big(\frac{T_{sem}}{T} \Big)^2} ~~,~~~~T = \frac{1}{2\pi k_B}M_{cl} c^2$$
$T_{sem}$ is the semiclassical (Gibbons-Hawking) de Sitter temperature . In the limit $ M_{c\ell}\rightarrow M_{sem}$, which implies $ M_{cl} \rightarrow m_{Pl} $,  $S_{sem}(H)$ is dominated by
$$
S_{sem}(H)_{\Delta \rightarrow 0} =  -~k_B~ \ln ~\sqrt{1 - \frac{T_{sem}}{T}}~~+~O(1)
$$
This shows that a {\bf phase transition} takes place at $T \rightarrow T_{sem}$. This implies that the transition occurs for $M_{cl} \rightarrow m_{Pl}$, ie $T \rightarrow t_{Pl}$, (that is for high $H \rightarrow c/\ell_{Pl}$).  
This is a {\bf gravitational} like transition, similar to the de Sitter string transition we analysed above: the signature of this transition is the square root {\bf branch point} behavior at the critical mass (temperature) {\bf analogous} to the thermal self-gravitating gas behavior of point particles, refs \cite {8}, \cite{9}, and to the string gas in de Sitter space, ref.\cite {7}.  This is also the same behaviour as that found for the entropy of the Kerr black hole in the high angular momentum $ J\rightarrow M^2G/c$ regime, (extremal transition), ref.\cite{6}. This is {\bf universal}, in any number of space-time dimensions.


\end{document}